\pdfoutput=1 
\documentclass[conference,10pt,a4paper]{IEEEtran}
\IEEEoverridecommandlockouts
\usepackage{amsmath}
\usepackage{amsopn}

\usepackage{float}

\usepackage{amssymb}
\usepackage{pifont}
\usepackage{amsfonts}

\usepackage{amsthm}

\usepackage{graphicx}

\usepackage{booktabs}

\usepackage{caption3} 
\DeclareCaptionOption{parskip}[]{} 
\usepackage[small]{caption}

\usepackage{setspace,graphicx,multirow}
\usepackage{subcaption}

\usepackage{algorithm}
\usepackage[]{algpseudocode}
\makeatletter
\def\BState{\State\hskip-\ALG@thistlm}
\makeatother
\usepackage{algcompatible}

\usepackage{float}

\usepackage{makecell}
\usepackage{geometry}
\usepackage{cite}

\usepackage{fancyhdr} 

\usepackage{color}
\ifodd 1
\newcommand{\rev}[1]{{\color{red}#1}} 
\newcommand{\revBlu}[1]{{\color{blue}#1}} 
\newcommand{\com}[1]{\textbf{\color{blue} (COMMENT: #1)}} 
\else
\newcommand{\rev}[1]{#1}
\newcommand{\revBlu}[1]{#1}

\newcommand{\com}[1]{}
\fi

\usepackage{enumitem}
\setenumerate[1]{itemsep=0pt,partopsep=0pt,parsep=\parskip,topsep=0pt}
\setitemize[1]{itemsep=0pt,partopsep=0pt,parsep=\parskip,topsep=0pt}
\setdescription{itemsep=0pt,partopsep=0pt,parsep=\parskip,topsep=0pt}

\usepackage[update,prepend]{epstopdf}

\begin{document}
\bibliographystyle{IEEEtran}
\bstctlcite{IEEEexample:BSTcontrol}

\title{AFDM Channel Estimation in\\Multi-Scale Multi-Lag Channels}

 \author{\IEEEauthorblockN{Rongyou Cao,~\IEEEmembership{Student Member,~IEEE}, Yuheng Zhong,~\IEEEmembership{Student Member,~IEEE}, Jiangbin Lyu,~\IEEEmembership{Member,~IEEE},\\
 Deqing Wang,~\IEEEmembership{Member,~IEEE}, and Liqun Fu,~\IEEEmembership{Senior Member,~IEEE}}

\thanks{This work was supported in part by the Natural Science Fundation of Xiamen (3502Z202372002), the Natural Science Foundation of Fujian Province (2023J01002), the Sichuan Science and Technology Program (2024NSFSC1416), the Guangdong Basic and Applied Basic Research Foundation (2023A1515030216), the National Natural Science Foundation of China (U23A20281, 62271427), the Key Science and Technology Project of Fujian Province (2023H0001), and the Fundamental Research Funds for the Central Universities (20720220078). The authors are with the School of Informatics, Xiamen University, China, the Sichuan Institute of Xiamen University, China, and the Shenzhen Research Institute of Xiamen University, China. \textit{Corresponding author: Jiangbin Lyu}, email: ljb@xmu.edu.cn.}%
}
\maketitle


\begin{abstract}
Affine Frequency Division Multiplexing (AFDM) is a brand new chirp-based multi-carrier (MC) waveform for high mobility communications, with promising advantages over Orthogonal Frequency Division Multiplexing (OFDM) and other MC waveforms. Existing AFDM research focuses on wireless communication at high carrier frequency (CF), which typically considers only Doppler frequency shift (DFS) as a result of mobility, while ignoring the accompanied Doppler time scaling (DTS) on waveform. However, for underwater acoustic (UWA) communication at much lower CF and propagating at speed of sound, the DTS effect could not be ignored and poses significant challenges for channel estimation. This paper analyzes the channel frequency response (CFR) of AFDM under multi-scale multi-lag (MSML) channels, where each propagating path could have different delay and DFS/DTS. Based on the newly derived input-output formula and its characteristics, two new channel estimation methods are proposed, i.e., AFDM with iterative multi-index (AFDM-IMI) estimation under low to moderate DTS, and AFDM with orthogonal matching pursuit (AFDM-OMP) estimation under high DTS. Numerical results confirm the effectiveness of the proposed methods against the original AFDM channel estimation method. Moreover, the resulted AFDM system outperforms OFDM as well as Orthogonal Chirp Division Multiplexing (OCDM) in terms of channel estimation accuracy and bit error rate (BER), which is consistent with our theoretical analysis based on CFR overlap probability (COP), mutual incoherent property (MIP) and channel diversity gain under MSML channels.
\end{abstract}


\section{Introduction}
Next-generation wireless systems are envisioned to support a wide range of services including communication in high mobility scenarios and in complex environments (e.g., underwater). This calls for new waveform design to cope with various highly demanding requirements \cite{bemani2021afdm}. Existing waveforms, such as Orthogonal Frequency Division Multiplexing (OFDM), have proved to achieve satisfactory or even optimal performance in time-invariant frequency selective channels. Nevertheless, for high mobility scenarios, OFDM does not achieve full diversity in linear time-varying (LTV) channels, thus unable to achieve optimal performance\cite{hadani2017orthogonal}. To address this issue, Affine Frequency Division Multiplexing (AFDM) is proposed in \cite{bemani2021afdm}, a new multi-carrier (MC) waveform using multiple orthogonal chirp signals to carry information.
Compared with OFDM and another chirp-based MC waveform, i.e., Orthogonal Chirp Division Multiplexing (OCDM)\cite{7523229}, AFDM has more adjustable parameters, allowing the multipaths to be separated from each other in the discrete affine Fourier transform (DAFT) domain\footnote{We refer such a property as the \textit{multi-path separability (MPS)}.}. In fact, AFDM includes OFDM and OCDM as special cases and is compatible with the standard OFDM procedure\cite{bemani2023affine}.

The research on AFDM is still in the early stage, which yet suggests AFDM to be a promising new waveform for high mobility scenarios\cite{bemani2023affine,Savaux_2023,tang2023time,yin2022pilot,benzine2023affine,ni2022afdmbased,Bemani_2024}.
The authors in \cite{Savaux_2023} deal with discrete Fourier transform (DFT)-based (de)modulation techniques for AFDM instead of DAFT, such that AFDM can be interpreted as a precoded OFDM waveform compliant with existing systems. 
The authors in \cite{tang2023time} investigate the synchronization of AFDM and propose the maximum-likelihood (ML) criteria by exploiting the redundant information contained within the chirp-periodic prefix (CPP) inherent in AFDM symbols.
The authors in \cite{yin2022pilot} propose a pilot aided channel estimation for AFDM, which achieves similar performance compared to AFDM with ideal channel state information (CSI).
The authors in \cite{benzine2023affine} investigate the minimal pilot and guard overhead needed for achieving a target mean squared error (MSE) when performing channel estimation for different waveforms on sparse time-varying channels, which suggests AFDM to be more efficient compared with single-carrier modulation (SCM), OFDM and Orthogonal Time Frequency Space (OTFS)\cite{hadani2017orthogonal}.
An AFDM-based integrated sensing and communications (ISAC) system is considered in \cite{ni2022afdmbased}, which decouples delay and Doppler shift in the fast time axis and maintains good sensing performance even in large Doppler shift scenarios. Further research \cite{Bemani_2024} indicates that ISAC systems based on OTFS, OCDM, and AFDM have comparable performance in terms of range and velocity estimation MSE, whereas the former two schemes require complex successive interference cancellation (SIC) while AFDM requires only a simple receiver architecture.

Despite the exciting research on AFDM and their promising results, whether and how AFDM can be extended to more complex channel environments remain largely as open questions.
This paper makes an attempt to bring AFDM to underwater acoustic (UWA) channels, 
which are known to be much more complex than radio channels\cite{li2007estimation}.
The UWA channels are typically modelled as wideband LTV channels with some fundamental differences against those experienced in radio systems such as cellular and WiFi.
Specifically, since the UWA channel has much lower carrier frequency (CF) and propagating at speed of sound, the Doppler effect due to mobility cannot be approximated by Doppler frequency shift (DFS) only. Rather, signals are compressed or dilated measurably due to \textit{Doppler time scaling (DTS)}. Furthermore, due to distinct angles of arrival in a multipath environment, each path might experience different DFS/DTS and/or delay. These effects give rise to the multi-scale multi-lag (MSML) channel model\cite{berger2009sparse}. 
Doppler factor, time delay, and complex amplitude of each multipath are the parameters to be estimated from the received signal, which is typically sampled with an identical sampling rate. Then using resampling\cite{berger2009sparse}, a single Doppler scale could be compensated, which yet leaves residual sampling errors to other scaled components and requires additional inter-carrier interference (ICI) suppression\cite{7494962}.
On the other hand, ML-based correlation methods could be applied for the MSML channel estimation, which yet requires solving a multi-dimensional non-linear least-squares (LS) problem, incurring a high complexity\cite{7131580}. 

Inspired by the appealing advantage of AFDM, this paper attempts to address the challenges in UWA MSML channels by employing the new AFDM waveform.
However, the original AFDM embedded channel estimation (AFDM-ECE) method\cite{bemani2023affine} may not be applicable or may result in poor accuracy, since the aforementioned MPS property of AFDM is compromised or even jeopardised depending on the level of DTS in MSML channels, which spreads the energy of each path to other paths and causes ambiguity when estimating their channel parameters.
To this end, we first analyze the channel frequency response (CFR) of AFDM under MSML channels. Based on the newly derived input-output formula and its characteristics, two new channel estimation methods are proposed, i.e., AFDM with iterative multi-index (AFDM-IMI) estimation under low to moderate DTS, and AFDM with orthogonal matching pursuit (AFDM-OMP) estimation under high DTS. Numerical results confirm the effectiveness of the proposed methods against the original AFDM-ECE method. Moreover, the resulted AFDM system outperforms OFDM as well as OCDM in terms of channel estimation accuracy and bit error rate (BER), which is consistent with our theoretical analysis based on CFR overlap probability (COP), mutual incoherent property (MIP) and
channel diversity gain under MSML channels.

\textit{Notations:} Boldface uppercase (lowercase) letters are used for matrices (vectors). The notations $\left(\cdot\right)^H$, $\left(\cdot\right)^T$ and $\left(\cdot\right)^{-1}$ denote the Hermitian transpose, transpose and matrix inverse, respectively. $I_K$ indicates a $K\times K$ identity matrix and ${\bf{0}}_{1\times K}$ denotes a $1\times K$ zero vector. The notation $\left(\cdot\right)_N$ indicates the modulo $N$ operation. The notation $\text{Sa}\left(\cdot\right)=\mathrm{sin}\left(\cdot\right)/\left(\cdot\right)$ denotes the sampling function, while $\delta\left(\cdot\right)$ denotes the Dirac delta function. The notation $\langle \mathbf{x},\mathbf{y}\rangle$ denotes the inner product of vector 
$\mathbf{x}$ and vector $\mathbf{y}$. The notation $\|\cdot\|$ denotes the Euclidean norm. The circularly symmetric complex Gaussian (CSCG) distribution with variance $\sigma^2$ is denoted as $\mathcal{CN}(0,\sigma^2)$.

\section{AFDM Input/Output Relations and Diversity Analysis in MSML Channels}\label{SectionDesign}
\begin{figure}
        \centering
        \begin{subfigure}{0.25\linewidth}
            \centerline{\includegraphics[width=0.95\linewidth, trim=15 5 10 20,clip]{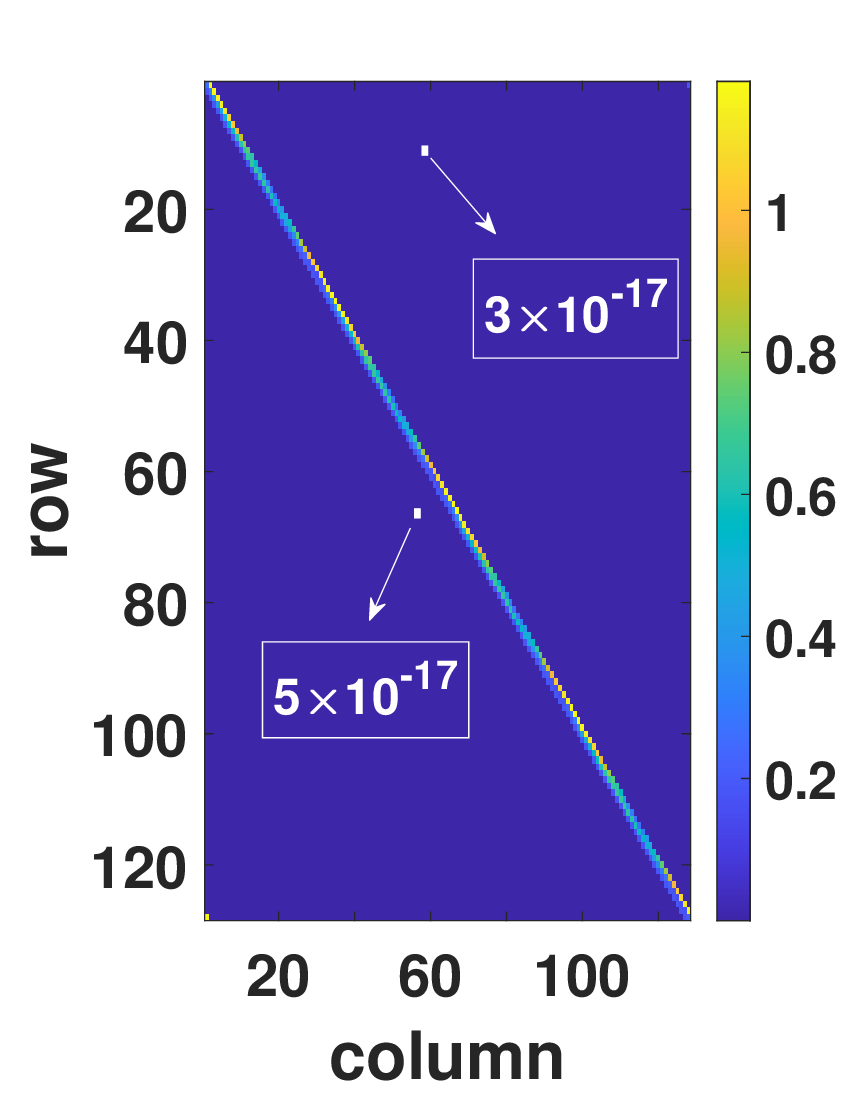}}
                \caption{}
                \label{Fig_OFDM_CFR_LTV}
        \end{subfigure}%
        \centering    
        \begin{subfigure}{0.25\linewidth}
            \centerline{\includegraphics[width=0.95\linewidth, trim=15 5 10 20,clip]{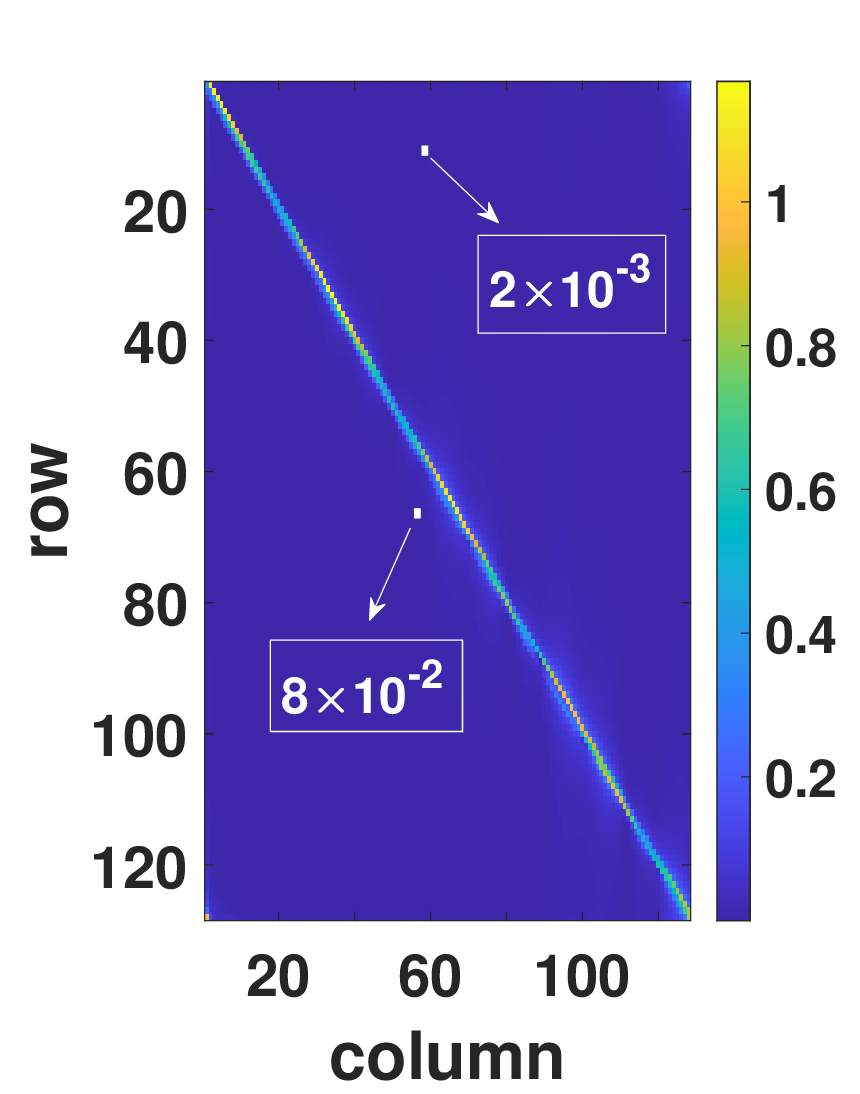}}
                \caption{}
                \label{Fig_OFDM_CFR_MSML}
        \end{subfigure}%
                \centering
        \begin{subfigure}{0.25\linewidth}
            \centerline{\includegraphics[width=0.95\linewidth, trim=15 5 10 20,clip]{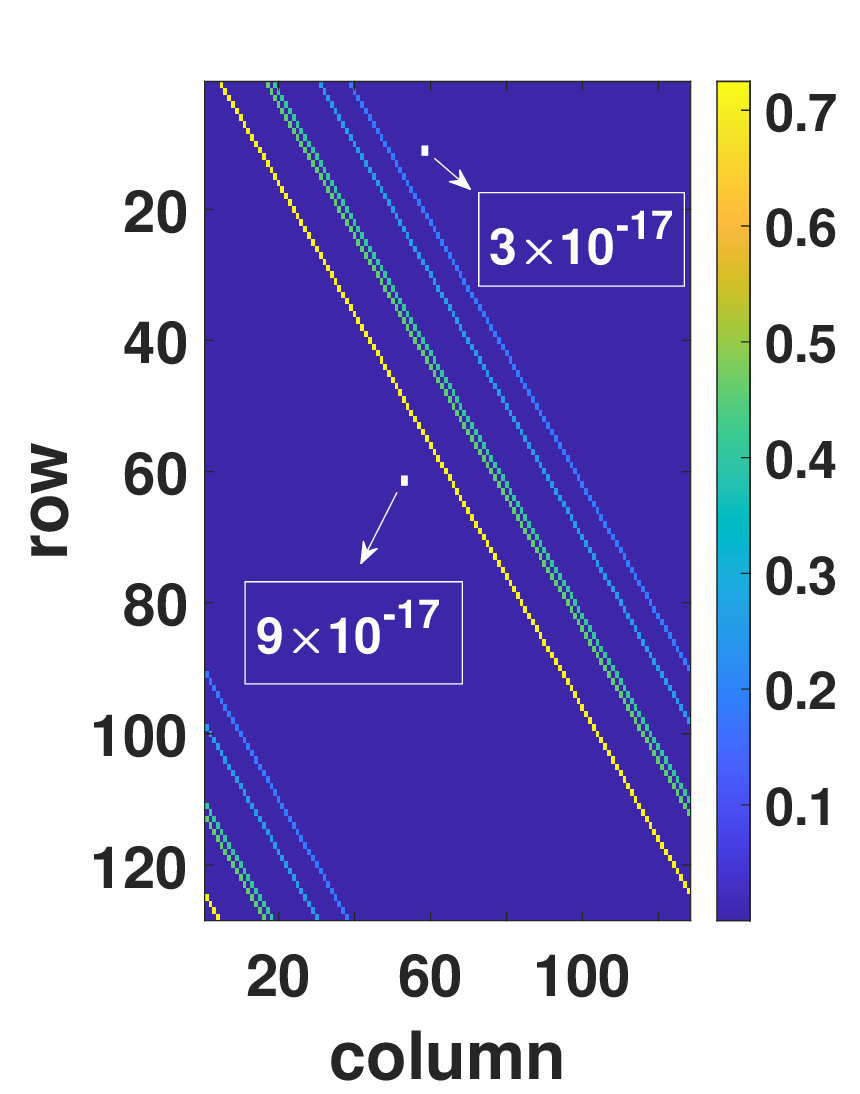}}
                \caption{}
                        \label{Fig_AFDM_CFR_LTV}
        \end{subfigure}%
        \centering    
        \begin{subfigure}{0.25\linewidth}
            \centerline{\includegraphics[width=0.95\linewidth, trim=15 5 10 20,clip]{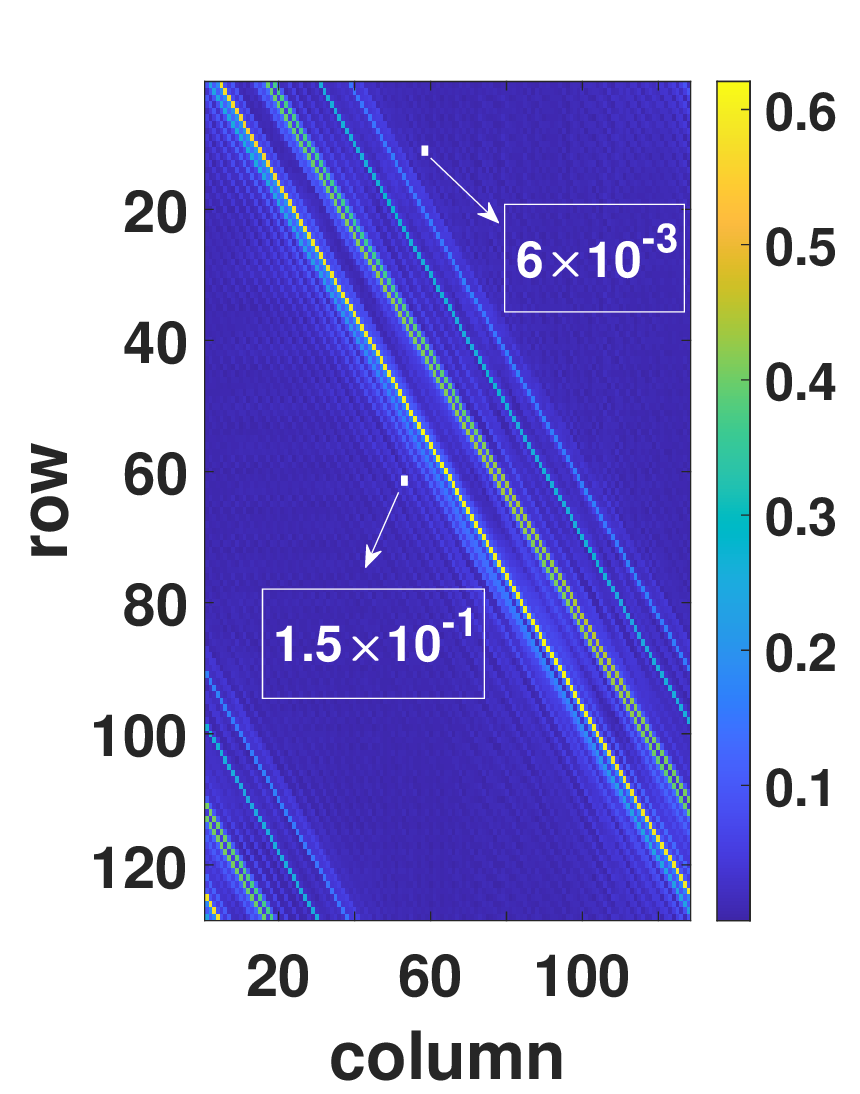}}
                \caption{}
                        \label{Fig_AFDM_CFR_MSML}
        \end{subfigure}%
        \caption{The per-entry amplitude of the CFR matrix $\mathbf{\bar{H}}$ of: (a) OFDM in non-scaled LTV channel; (b) OFDM in MSML channel; (c) AFDM in non-scaled LTV channel; and (d) AFDM in MSML channel under $P=5$, $l_\textrm{max}=19$, $Q_\textrm{max}=1$ and Doppler factor in order of $10^{-3}$.\vspace{-3ex}}
    \label{Fig_CFR}
\end{figure}

In this section, we introduce the MSML channel model and derive the CFR of AFDM in it. Comparison of the CFRs of OFDM and AFDM is illustrated in Fig. \ref{Fig_CFR}, with detailed derivations in the following.

\subsection{Transmitter}
Let $\mathbf{s}\in \mathbb{A}^{N\times1}$ denote the vector of information symbols in the DAFT domain, where $N$ and $\mathbb{A}$ represent the total numbers of subcarriers (SCs) and the complex alphabet, respectively. The AFDM-modulated discrete time-domain signal $\mathbf{x}$ is\cite{bemani2023affine}
\begin{equation}\label{modulated_x}
x\left[n\right]\triangleq\sum\nolimits_{m=0}^{N-1}s\left[m\right]\phi_n\left[m\right],\quad n=0,\cdots,N-1,
\end{equation}
where $\phi_n\left[m\right]\triangleq\frac1{\sqrt{N}}\cdot e^{j2\pi\left(c_1n^2+c_2m^2+nm/N\right)}$ represents the $\left(m,n\right)$-th entry of the inverse DAFT (IDAFT) matrix $\mathbf{A}^H$ ($\mathbf{A}$ represents the DAFT matrix with parameters $c_1$ and $c_2$). Note that when $c_1=c_2=0$, $\phi_n\left[m\right]=\frac1{\sqrt{N}}\cdot e^{j2\pi nm/N}$ turns into the $\left(m,n\right)$-th entry of the inverse DFT (IDFT) matrix $\mathbf{F}^H$ ($\mathbf{F}$ represents the DFT matrix). In matrix form, (\ref{modulated_x}) becomes $\mathbf{x}=\mathbf{A}^H\mathbf{s}=\mathbf{\Lambda}_{c_1}^H \mathbf{F}^H\mathbf{\Lambda}_{c_2}^H \mathbf{s}$, where $\mathbf{\Lambda}_{c}$ is a diagonal matrix with the $n$-th diagonal entry being $e^{-j2\pi cn^2}$, which adds only marginal increases in complexity and memory compared with OFDM.
Consider an AFDM system with a bandwidth of $f_\mathrm{s}$ Hz and symbol length of $T$ seconds, of which $N_\mathrm{CPP}$ is the length of the chirp-periodic prefix (CPP)\cite{bemani2023affine} that is usually no shorter than the value in samples of the maximum delay spread $\tau_\textrm{max}$ of the channel, i.e., $l_\textrm{max}\triangleq\tau_\textrm{max}f_\mathrm{s}$. After inserting the CPP,\footnote{In this paper $2Nc_1$ is an integer value and $N$ is even, so the CPP is simply the cyclic prefix (CP)\cite{benzine2023affine}.} $\mathbf{x}$ is then converted to a serial baseband time-domain signal $X\left(t\right)$ through a Digital/Analog (D/A) converter (i.e., convolution with the sampling function), given by
\begin{equation}\label{x_1}
    X\left(t\right)\triangleq\sum_{n=-N_\mathrm{CPP}}^{N-1}x\left[\left(n\right)_N\right]\cdot \text{Sa}\left(\pi f_\mathrm{s}\left(t-\frac{n}{f_\mathrm{s}}\right)\right).
\end{equation}

\subsection{MSML Channel}
The channel impulse response (CIR) of an MSML channel for passband is given by \cite{li2007estimation}
\begin{equation}
    h\left(t,\tau\right)\triangleq\sum\nolimits_{i=1}^{P}h_i\cdot\delta\left(\tau-\tau_i+\rev{a_it}\right)\cdot \revBlu{e^{j2\pi a_if_\mathrm{c}t}},
\end{equation}
where $P$, $f_\mathrm{c}$, $h_i$, $a_i$, $\tau_i$ are the number of paths, the carrier frequency, the complex gain, the Doppler factor, and the delay associated with the $i$-th path, respectively. After transmission over the channel, the serial received signal is given by
\begin{equation}
\begin{aligned}
    Y\left(t\right)&\triangleq X\left(t\right)*h\left(t,\tau\right)+w\left(t\right)\\
    &=\sum\nolimits_{i=1}^{P}h_iX\big(\left(1+\rev{a_i}\right)t-\tau_i\big)\revBlu{e^{j2\pi a_if_\mathrm{c}t}} + w\left(t\right),
\end{aligned}
\end{equation}
where $w\left(t\right)$ denotes the additive white Gaussian noise (AWGN). 
Note that in radio systems the DTS term $\left(1+a_i\right)t$ is typically approximated as $t$ only, while only considering the DFS term $e^{j2\pi a_if_\mathrm{c}t}$. However, the DTS cannot be ignored in UWA channels, as discussed in Section I.

\subsection{Receiver}
Through a Analog/Digital (A/D) converter with a sampling rate of $f_\mathrm{s}$, the discrete time-domain received signal $\mathbf{y}$ is obtained by removing the CPP, with the $m$-th entry given by
\begin{equation}\label{ym_1}
\begin{aligned}
    &y\left[m\right]\triangleq Y(mT/N)=Y(m/f_\mathrm{s})\\
    &=\sum\nolimits_{i=1}^{P}h_iX\big(\left(1+a_i\right)\frac{m}{f_\mathrm{s}}-\tau_i\big)e^{j2\pi a_if_\mathrm{c}\frac{m}{f_\mathrm{s}}}+w\big(\frac{m}{f_\mathrm{s}}\big).
\end{aligned}
\end{equation}
Combined with (\ref{x_1}), eq.(\ref{ym_1}) can be rewritten as
\begin{small}
\begin{align}    
    y\left[m\right]=&\sum_{i=1}^{P}h_i\sum_{n=-N_\mathrm{CPP}}^{N-1}x\left[\left(n\right)_N\right]\cdot \text{Sa}\big(\pi((1+a_i)m-f_\mathrm{s}\tau_i-n)\big)\notag\\
    &\times e^{j2\pi D_im}+w\big(\frac{m}{f_\mathrm{s}}\big),
\end{align}
\end{small}%
where $m\in\mathcal{N}\triangleq\{0,1,\cdots,N-1\}$, and $D_i\triangleq a_i\cdot f_\mathrm{c}/f_\mathrm{s}$ denotes the DFS (in digital frequencies) associated with the $i$-th path. In matrix form, $\mathbf{y}$ is related to $\mathbf{x}$ by
\begin{equation}
\mathbf{y}=\sum\nolimits_{i=1}^{P}h_i\mathbf{H}_i\mathbf{x}+\mathbf{w}=\mathbf{Hx}+\mathbf{w},
\end{equation}
where $\mathbf{w}$ denotes the discrete AWGN vector, comprising $w\left[m\right]\triangleq w\left(t\right)|_{t=\frac{m}{f_\mathrm{s}}}$, $m\in\mathcal{N}$. The time-domain channel matrix $\mathbf{H}$ is defined as $\mathbf{H}\triangleq\sum_{i=1}^{P}h_i\mathbf{H}_i$, where $\mathbf{H}_i$ is an $N\times N$ matrix for the $i$-th path whose $\left(m,n\right)$-th entry represents the impact of input $x\left[n\right]$ on the output $y\left[m\right]$, as given by
\begin{equation}
    {\small H_i[m,n]=}{\scriptsize\left\{
        \begin{array}{cc}
        g_{m,n}e^{j2\pi D_im},& n\in\left[0,N-N_\mathrm{CPP}-1\right],\\
        \left(g_{m,n}+g_{m,\left(n-N\right)}\right)e^{j2\pi D_im},& n\in\left[N-N_\mathrm{CPP},N-1\right].\label{Hi}
        \end{array}
    \right.}
\end{equation}
where $g_{m,n}=\text{Sa}\left(\pi f_{m,n}\right)$ and $f_{m,n}=\left(1+a_i\right)m-f_\mathrm{s}\tau_i-n$. Note that DTS is neglected in radio channels, and hence for any $n$, there is only one unique $m$ making $g_{m,n}$ and hence $H_i[m,n]$ non-zero (i.e., when $m=f_\mathrm{s}\tau_i+n$), thus making $H_i$ a pseudo-cyclic matrix. 
However, DTS cannot be ignored in UWA channels, and hence $H_i[m,n]$ is non-zero for any $m$ or $n$, causing the energy of each path to spread to positions other than $m=f_\mathrm{s}\tau_i+n$. This leads to \textit{ambiguity} in distinguishing between different multipaths.

Finally, the AFDM demodulated information vector $\mathbf{z}$ is obtained by multiplying $\mathbf{y}$ with the DAFT matrix $\mathbf{A}$, i.e.,
\begin{equation}
    \mathbf{z}\triangleq\mathbf{Ay}.
\end{equation}


\subsection{Input-Output Relations}\label{InputOutput}
In matrix representation, the output can be written as
\begin{small}
\begin{equation}\label{zyz}
\mathbf{z}=\mathbf{Ay}=\mathbf{AHx}+\mathbf{Aw}=\mathbf{AHA}^H\mathbf{s}+\mathbf{\bar{w}}=\mathbf{\bar{H}s}+\mathbf{\bar{w}},%
\end{equation}%
\end{small}%
where $\mathbf{\bar{H}}\triangleq \mathbf{AHA}^H$ and $\mathbf{\bar{w}}=\mathbf{Aw}$ represent the DAFT-domain CFR matrix and noise vector, respectively. Since $\mathbf{A}$ is a unitary matrix, $\mathbf{\bar{w}}$ and $\mathbf{w}$ have the same covariance.
Since $\mathbf{H}\triangleq\sum_{i=1}^{P}h_i\mathbf{H}_i$, we have
\begin{small}
\begin{equation}\label{Hi_definition}
\mathbf{\bar{H}}=\sum\nolimits_{i=1}^{P}h_i\mathbf{\bar{H}}_i=\sum\nolimits_{i=1}^{P}h_i\mathbf{\Lambda}_{c_2}\mathbf{F\Lambda}_{c_1}\mathbf{H}_i\mathbf{\Lambda}_{c_1}^{H}\mathbf{F}^H\mathbf{\Lambda}_{c_2}^H,%
\end{equation}%
\end{small}%
where $\mathbf{\bar{H}}_i\triangleq\mathbf{\Lambda}_{c_2}\mathbf{F\Lambda}_{c_1}\mathbf{H}_i\mathbf{\Lambda}_{c_1}^{H}\mathbf{F}^H\mathbf{\Lambda}_{c_2}^H$ represents the CFR matrix of the $i$-th path with unit channel gain.

Define $\mathbf{B}_i\triangleq\mathbf{\Lambda}_{c_1}\mathbf{H}_i\mathbf{\Lambda}_{c_1}^H$, whose $\left(m,n\right)$-th entry is given by
\begin{small}
\begin{equation}
    B_i\left[m,n\right]=e^{j2\pi c_1\left(n^2-m^2\right)}H_i\left[m,n\right].
\end{equation}
\end{small}
Define $\mathbf{C}_i\triangleq \mathbf{B}_i\mathbf{F}^H$, whose $\left(m,n\right)$-th entry is given by
\begin{small}
\begin{equation}
    C_i\left[m,n\right]=\sum\nolimits_{k=0}^{N-1}B_i\left[m,k\right]F^H\left[k,n\right].%
\end{equation}%
\end{small}%
Further denote $\eta\triangleq e^{j2\pi c_2\left(n^2-m^2\right)}$. Then from (\ref{Hi_definition}), we have $\mathbf{\bar{H}}_i=\mathbf{\Lambda}_{c2}\mathbf{FC}_i\mathbf{\Lambda}_{c2}^H$, whose $\left(m,n\right)$-th entry is given by
\begin{small}
\begin{align}
    &\bar{H}_i\left[m,n\right]=e^{j2\pi c_2\left(n^2-m^2\right)}\sum_{q=0}^{N-1}F\left[m,q\right]C_i\left[q,n\right]\notag\\
    &=\eta\sum_{q=0}^{N-1}\frac{1}{\sqrt{N}}e^{-j2\pi\frac{mq}{N}}\sum_{k=0}^{N-1}B_i\left[q,k\right]F^H\left[k,n\right]\notag\\
    &=\frac{\eta}{\sqrt{N}}\sum_{q=0}^{N-1}\left(e^{-j2\pi\frac{mq}{N}}\sum_{k=0}^{N-1}\frac{1}{\sqrt{N}}e^{j2\pi c_1\left(k^2-q^2\right)+j2\pi\frac{kn}{N}}H_i\left[q,k\right]\right)\notag\\
    &=\frac{\eta}{N}\sum_{q=0}^{N-1}\left(e^{-j2\pi\left(\frac{mq}{N}+c_1q^2\right)}\sum_{k=0}^{N-1}e^{j2\pi \big(\frac{kn}{N}+c_1k^2\big)}H_i\left[q,k\right]\right).\label{Hibar}
\end{align}
\end{small}

Define $Q_i\triangleq ND_i$ and $l_i\triangleq f_\mathrm{s}\tau_i$, where $Q_i\in\left[-Q_\textrm{max},Q_\textrm{max}\right]$ is the DFS normalized with respect to the SC spacing $f_\mathrm{s}/N$, and $l_i\in\left[0,l_\textrm{max}\right]$ is the delay normalized with respect to the sampling interval $1/f_\mathrm{s}$.\footnote{Assuming a maximum delay of $\tau_\mathrm{max}$ and a maximum Doppler factor of $a_\text{max}$, we have $l_\textrm{max}=f_\mathrm{s}\tau_\mathrm{max}$ and $Q_\textrm{max}=N a_\text{max}f_\mathrm{c}/f_\mathrm{s}$.} In this paper, $Q_i$ is assumed to be integer valued for all $i\in\left\{1,\cdots,P\right\}$, and $c_1$ is chosen such that $2Nc_1l_i$ is an integer\cite{bemani2023affine}. When DTS is small or ignored, the above configuration ensures that $\mathbf{\bar{H}}_i$ associated with the $i$-th path is effectively a pseudo-cyclic matrix, i.e., a cyclically shifted version of a diagonal matrix with non-zero entries at $n=\left(m+\textrm{loc}_i\right)_N$, where $\textrm{loc}_i\triangleq\left(Q_i+2Nc_1l_i\right)_N$.
In this paper, we propose a new performance indicator under non-scaled/scaled LTV channels, i.e., the \textit{CFR overlap probability (COP)} between two paths $i$ and $j$, defined as
\begin{equation}\label{Formula_COP}
\mathbb{P}(\textrm{loc}_i=\textrm{loc}_j)=\mathbb{P}\big((Q_i+2Nc_1l_i)_N=(Q_j+2Nc_1l_j)_N\big).
\end{equation}
When $\textrm{loc}_i=\textrm{loc}_j$, the two paths $i$ and $j$ will overlap in CFR, and hence are strongly correlated/difficult to be distinguished.
The COP depends on the random channel delays and DFSs, and can be controlled by the parameter $c_1$.
Note that AFDM includes OFDM ($c_1=0$) and OCDM ($c_1=1/(2N)$) as special cases, whereby AFDM sets $c_1\geqslant\left({2Q_\mathrm{max}+1}\right)/\left({2N}\right)$ such that channel paths with different delays or DFSs will become separated in the DAFT domain\cite{bemani2023affine}, resulting in the overall CFR matrix $\mathbf{\bar{H}}$ with the structure shown in Fig. \ref{Fig_AFDM_CFR_LTV}, hence manifesting the MPS property. In comparison, for OFDM, paths with different delays $l_i$ cannot not be separated (except when they have different DFSs $Q_i$), as shown in Fig. \ref{Fig_OFDM_CFR_LTV}, manifesting poor MPS.
The performance of OCDM lies in between AFDM and OFDM, and is not shown for brevity.


However, in MSML channels, the non-negligible DTS spreads the energy of each path to positions other than $n=\left(m+\textrm{loc}_i\right)_N$, and hence $\mathbf{\bar{H}}_i$ is no longer a pseudo-cyclic matrix. 
In other words, according to \eqref{Hi} and \eqref{Hibar}, all entries of $\mathbf{\bar{H}}_i$ are non-zero for both OFDM and AFDM, as shown in Fig. \ref{Fig_OFDM_CFR_MSML} and Fig. \ref{Fig_AFDM_CFR_MSML}. 
Nevertheless, AFDM still has better MPS than OFDM and OCDM, although the power of each path is spread gradually off the positions $n=\left(m+\textrm{loc}_i\right)_N$, depending on the level of Doppler factors.

Finally, once the CFR matrix $\mathbf{\bar{H}}$ is estimated (see later in Section \ref{ChannelEstimation}), a minimum mean square error (MMSE) receiver can be applied for data demodulation, i.e.,
\begin{equation}
\mathbf{\hat{s}}=\big(\mathbf{\bar{H}}^H\mathbf{\bar{H}}+\frac{1}{\textrm{SNR}}\mathbf{I}_N\big)^{-1}\mathbf{\bar{H}}^H\mathbf{z},
\end{equation}
where $\textrm{SNR}$ denotes the transmit signal-to-noise ratio.

\subsection{Diversity Analysis}\label{SectionDiversity}
In this part, we extend the diversity analysis in\cite{bemani2023affine} to MSML channels, assuming known and given CSI. Firstly, define
\begin{small}
\begin{equation}
\mathbf{M}\left(\mathbf{s}\right)\triangleq\left[\mathbf{\bar{H}}_1\mathbf{s}|\cdots|\mathbf{\bar{H}}_P\mathbf{s}\right]_{N\times P},
\end{equation}
\end{small}%
which represents the collection of received signal copies through each of the $P$ paths.
Define $\boldsymbol{\delta}\triangleq\boldsymbol{\delta}^{\left(m,n\right)}=\mathbf{s}_m-\mathbf{s}_n$ which represents the error when transmitting symbol $\mathbf{s}_m$ and deciding in favor of $\mathbf{s}_n$ at the receiver.
We can then define the \textit{error measurement matrix} as $\mathbf{R}\triangleq\mathbf{M}\left(\boldsymbol{\delta}\right)^H\mathbf{M}\left(\boldsymbol{\delta}\right)$,
whose $(i,j)$-th entry $R_{i,j}\triangleq\boldsymbol{\delta}^H\mathbf{\bar{H}}_i^H\mathbf{\bar{H}}_j\boldsymbol{\delta}$ represents the inner product of $\mathbf{\bar{H}}_i\boldsymbol{\delta}$ and $\mathbf{\bar{H}}_j\boldsymbol{\delta}$. 
Assuming that the transmit SNR is $1/N_0$ and that complex gains $h_i\sim \mathcal{CN}(0,1/P)$ are independent and identically distributed (i.i.d.), the average pairwise error probability (PEP) between $\mathbf{s}_m$ and $\mathbf{s}_n$ is upper bounded as\cite{bemani2021afdm}
\begin{small}
\begin{equation}\label{Formula_PEP}
    \mathbb{P}\left(\mathbf{s}_m\rightarrow \mathbf{s}_n\right)\leqslant\prod_{l=1}^{r}\frac{1}{1+\frac{\lambda_l^2}{4PN_0}}\triangleq \overline{\textrm{PEP}},
\end{equation}
\end{small}%
where $r$ and $\lambda_l^2$ denote the rank and the $l$-th eigenvalue of the matrix $\mathbf{R}$, respectively.
It can be seen that higher values of $r$ and $\lambda_l^2$ lead to lower PEP.
The diversity order is defined as
\begin{small}
\begin{equation}
    \rho \triangleq \min_{\boldsymbol{\delta}^{\left(m,n\right)}, m\neq n} \textrm{rank}\left(\mathbf{R}\right)\quad\leq P.
\end{equation}
\end{small}%
Due to the spread caused by MSML channels as discussed in Section \ref{InputOutput}, we extend the above definition by considering the \textit{effective rank} whereby only significant eigenvalues $\lambda_l^2\geq \epsilon$ (e.g., $\epsilon=0.1$) are counted towards the rank.

\begin{figure}
        \centering
        \begin{subfigure}{0.25\linewidth}
            \centerline{\includegraphics[width=0.95\linewidth, trim=8 0 20 20,clip]{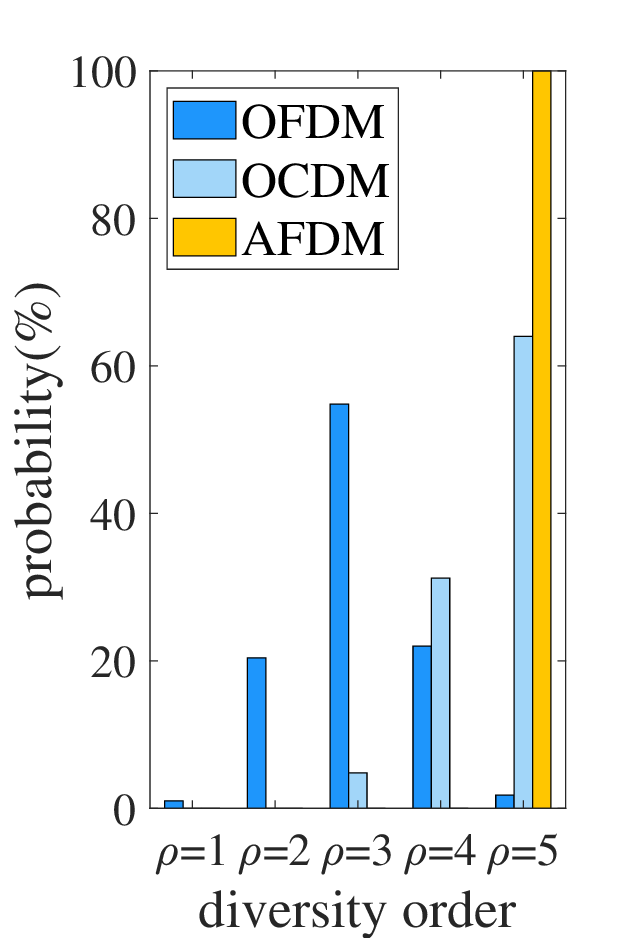}}
                \caption{}
                \label{Fig_DO_1e-4}
        \end{subfigure}%
        \centering    
        \begin{subfigure}{0.25\linewidth}
            \centerline{\includegraphics[width=0.95\linewidth, trim=8 0 20 20,clip]{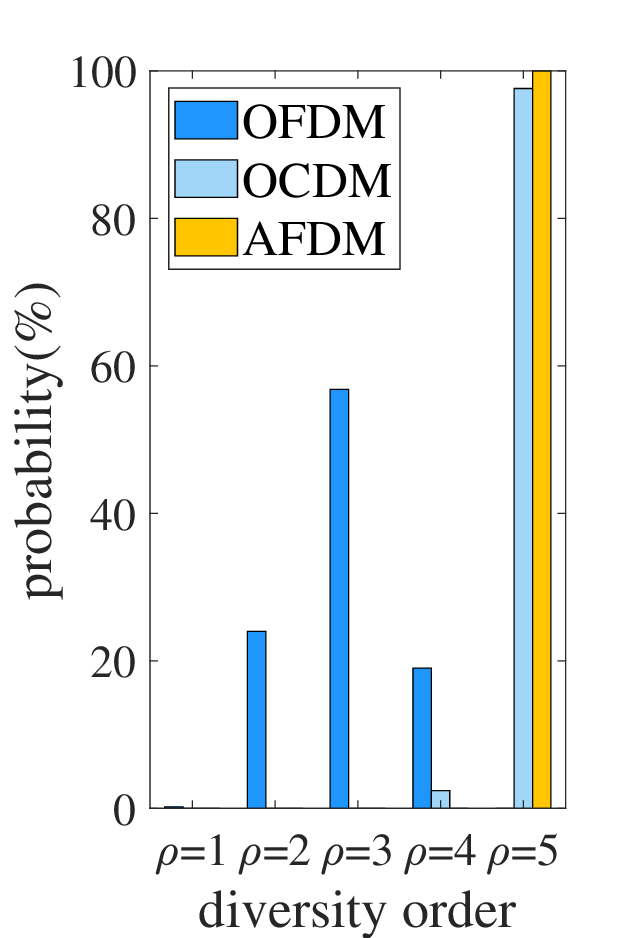}}
                \caption{}
                \label{Fig_DO_1e-2}
        \end{subfigure}%
                \centering
        \begin{subfigure}{0.25\linewidth}
            \centerline{\includegraphics[width=0.95\linewidth, trim=5 5 20 20,clip]{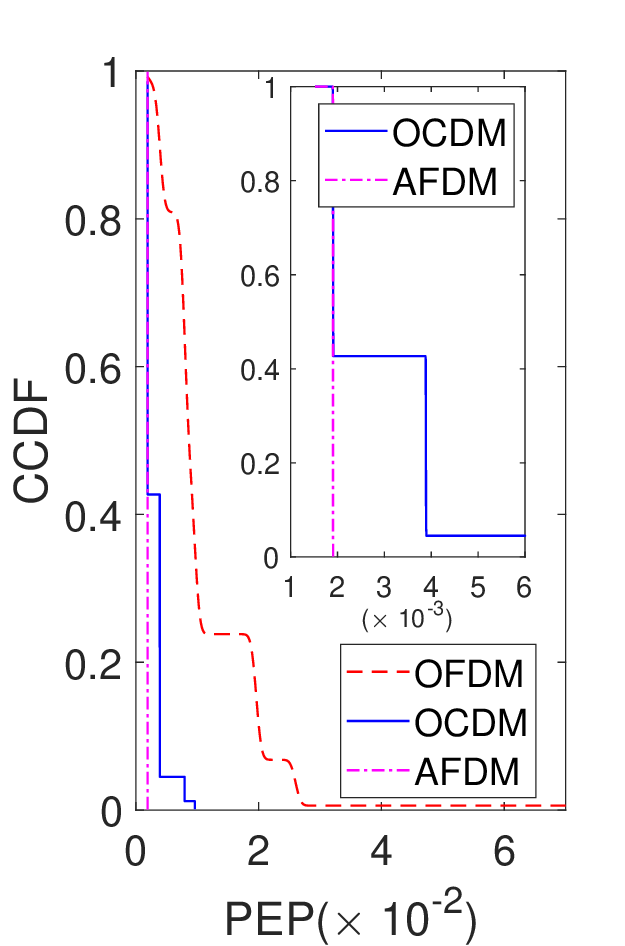}}
                \caption{}
                        \label{Fig_PEP_1e-4}
        \end{subfigure}%
        \centering    
        \begin{subfigure}{0.25\linewidth}
            \centerline{\includegraphics[width=0.95\linewidth, trim=5 5 20 20,clip]{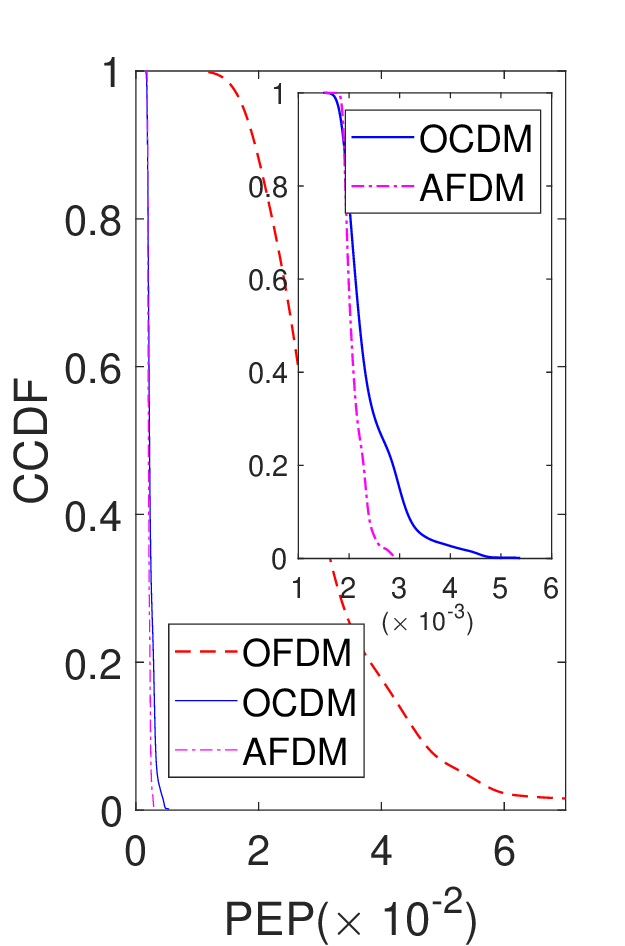}}
                \caption{}
                        \label{Fig_PEP_1e-2}
        \end{subfigure}%
        \caption{The occurrence probability of diversity order $\rho$ and the CCDF of PEP associated with OFDM, OCDM and AFDM, respectively, under Doppler factor in the order of (a) (c) $10^{-4}$ and (b) (d) $10^{-2}$.\vspace{-3ex}}
    \label{Fig_DO_PEP}
\end{figure}

Besides PEP and diversity order, our newly proposed COP indicator helps explaining the performance differences among AFDM, OCDM and OFDM.
When two paths overlap, i.e., $\textrm{loc}_i=\textrm{loc}_j$, the two paths $i$ and $j$ will overlap in CFR and hence highly correlated, which reduces the diversity order and increases the PEP.
For illustration, we plot the occurrence probability of diversity order $\rho$ and the complementary cumulative distribution function (CCDF) of $\overline{\textrm{PEP}}$ in \eqref{Formula_PEP} associated with OFDM, OCDM and AFDM, respectively, as shown in Fig. \ref{Fig_DO_PEP}, 
where the delay $l_i$ and DFS $Q_i$ are randomly drawn from uniform distribution with $l_\mathrm{max}=15$, $Q_\mathrm{max}=2$, $P=5$ and $1/N_0=50$.
In non-scaled LTV channels or when the DTS is small, it can be seen that $\rho_\textrm{OFDM}<\rho_\textrm{OCDM}<\rho_\textrm{AFDM}=P$, as shown in Fig. \ref{Fig_DO_1e-4}.
This is consistent with our analysis in Section \ref{InputOutput}, where AFDM has the best MPS while OFDM has the worst.
Consequently, AFDM has overall the lowest PEP, as shown in Fig. \ref{Fig_PEP_1e-4}.
On the other hand, in MSML channels, the non-negligible DTS spreads the power of each path $i$ gradually off the positions $n=\left(m+\textrm{loc}_i\right)_N$, causing all entries of $\mathbf{\bar{H}}_i$ to be non-zero.
For the above example setup, the impact on the diversity order and PEP is illustrated in Fig. \ref{Fig_DO_1e-2} and Fig. \ref{Fig_PEP_1e-2}, respectively.
For OFDM, the diversity order is not changed significantly, while its PEP performance becomes poorer.
For OCDM, it is interesting to observe that the diversity order is increased while its PEP performance is slightly improved.
For AFDM, it still achieves full diversity order and has robust PEP performance.
Finally, note that the above performance is achievable under known and given CSI, which thus calls for accurate channel estimation.

\section{AFDM Channel Estimation in MSML Channels}\label{ChannelEstimation}
In this section, we design two new channel estimation schemes for AFDM based on the derived CFR in MSML channels. 
The AFDM-IMI scheme has lower complexity and works well under low to moderate Doppler factors, while the AFDM-OMP scheme works also under high Doppler factors at the cost of slightly increased computational complexity.

\subsection{AFDM with Iterative Multi-Index (IMI) Estimation}
\begin{figure}
        \centering
        \begin{subfigure}{0.55\linewidth}
                \centerline{\includegraphics[width=1\linewidth,  trim=0 0 0 0,clip]{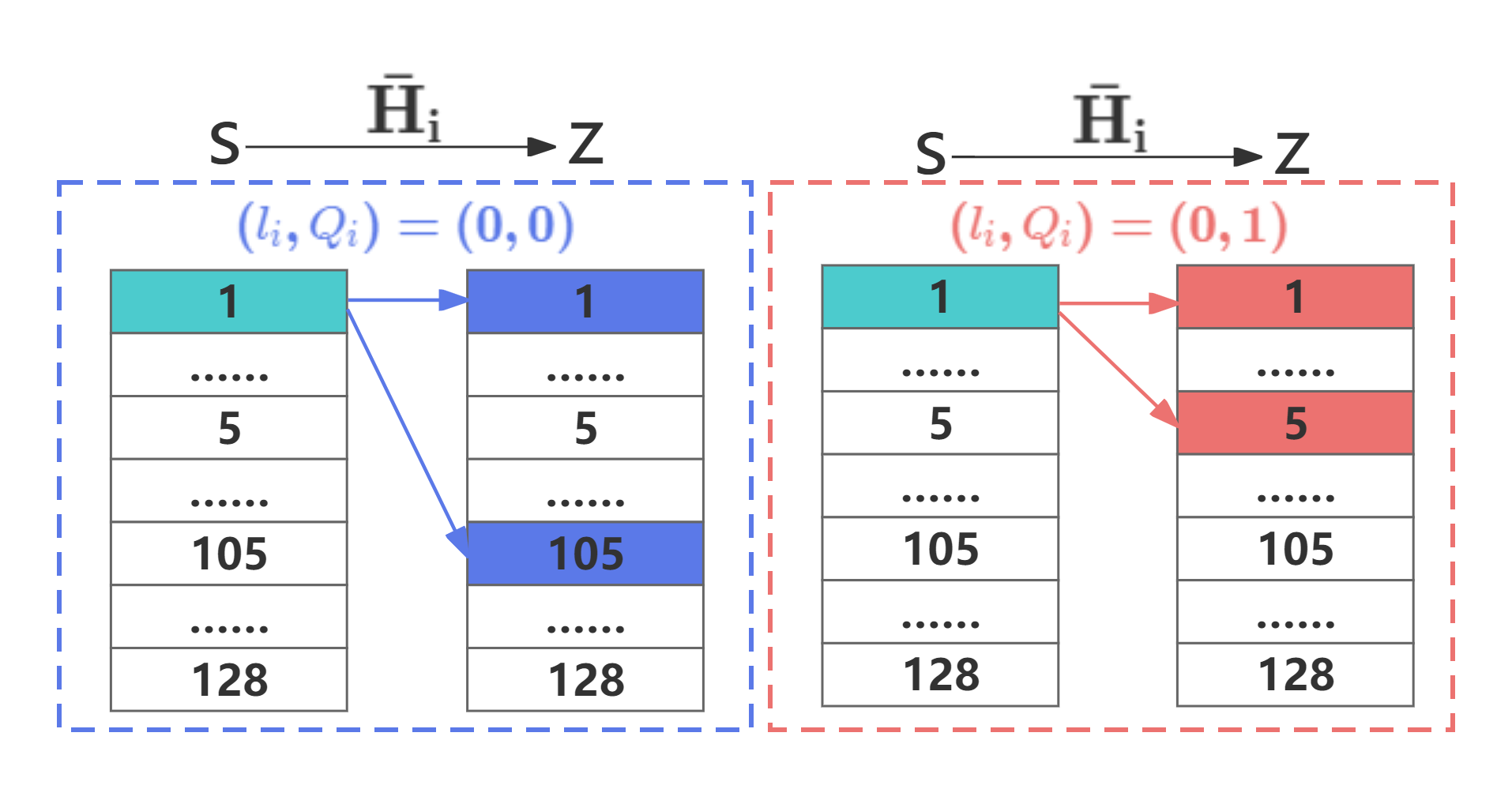}}
                \caption{}
                \label{Fig_index_positon}
        \end{subfigure}%
        \centering    
        \begin{subfigure}{0.45\linewidth}
                \centerline{\includegraphics[width=0.85\linewidth,  trim=0 0 0 0,clip]{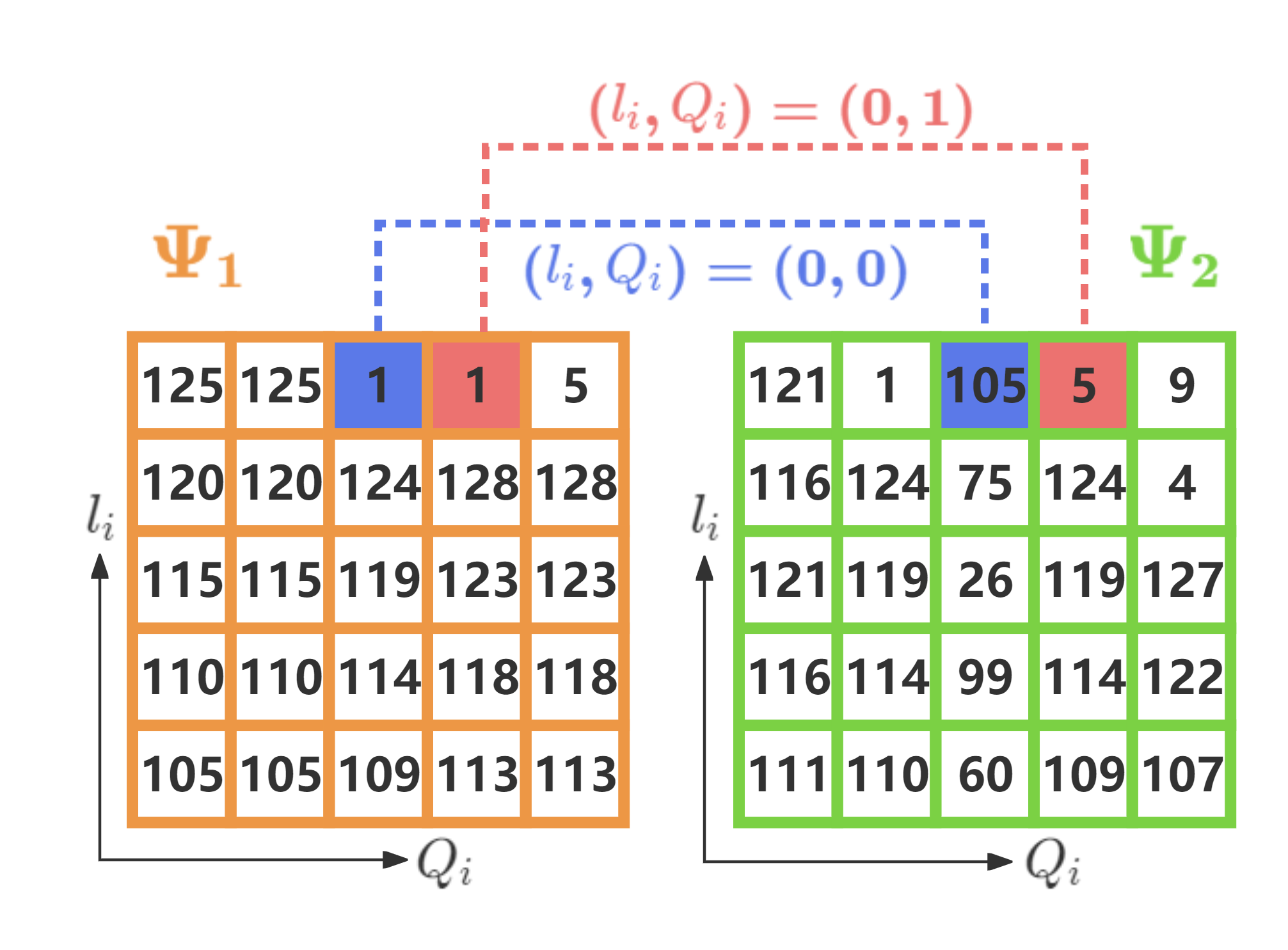}}
                \caption{}
                \label{Fig_MI_dictionary}
        \end{subfigure}%
        \caption{(a) Output $\mathbf{z}$ for the same input $\mathbf{s}$ under different $\left(l_i,Q_i\right)$ combination; and (b) the corresponding matrices $\mathbf{\Psi}_1$ and $\mathbf{\Psi}_2$.\vspace{-2ex}}\label{Fig_dictionary}
\end{figure}

Under small Doppler factors, the magnitude of $\bar{H}_i\left[m,n\right]$ has a peak at locations $n=\left(m+\textrm{loc}_i\right)_N$, which then generally decreases as $n$ moves away from $\left(m+\textrm{loc}_i\right)_N$, as shown in Fig. \ref{Fig_AFDM_CFR_MSML}. 
Taking the first column of $\mathbf{\bar{H}}_i$ as an example (i.e., $n=1$), the position $m$ where the peak appears is uniquely determined by the parameter $\textrm{loc}_i$, which in turn depends on $\left(l_i,Q_i\right)$. For different $\left(l_i,Q_i\right)$, we can thus obtain the index $m$ of the peak position
and store it in a matrix $\mathbf{\Psi}_1$, as shown in Fig. \ref{Fig_dictionary}.
Furthermore, based on our derived input/output formula for AFDM under MSML channels, we can also obtain the index $m$ of the second-peak position and store it in a matrix $\mathbf{\Psi}_2$, and so on.
These index matrices could then jointly help deciding the accurate $\left(l_i,Q_i\right)$ combination.

More specifically, consider the example where the pilot is inserted in the first SC of the information symbol $\mathbf{s}$, as shown in Fig. \ref{Fig_dictionary}.
The original AFDM-ECE method works under small Doppler factors where different $\left(l_i,Q_i\right)$ combination leads to different peak positions in the received symbol $\mathbf{z}$.
However, as the Doppler factor increases, it could happen that the peak position of $\bar{H}_i\left[m,n\right]$ no longer satisfies $n=\left(m+\textrm{loc}_i\right)_N$. In other words, the elements of the matrix $\mathbf{\Psi}_1$ may not be unique any more, as shown in Fig. \ref{Fig_dictionary}, where the $\left(l_i,Q_i\right)$ combination of (0,0) can not be distinguished from (0,1), rendering AFDM-ECE inapplicable.
To address the above issues, we propose the AFDM-IMI method which works by recursively looking at the index matrices $\mathbf{\Psi}_1$, $\mathbf{\Psi}_2$, and so on.
Specifically, in each iteration $k$, we aim to estimate the path with the $k$-th largest power gain, whose contribution is then eliminated from $\mathbf{z}$ after estimation of path parameters.
Take the first iteration for example.
The element index in $\mathbf{z}$ with the largest amplitude is denoted as $\omega$. 
Find the index set $\Omega$ such that $\mathbf{\Psi}_1(p,q)=\omega$, $\forall (p,q)\in \Omega$.
If $\Omega$ has only one element, then the corresponding $(l,Q)$ combination is uniquely determined.
Otherwise, say, there exist two elements in $\Omega$, denoted by $(p_1,q_1)$ and $(p_2,q_2)$.
We further distinguish between them by comparing the amplitude of $\mathbf{z}\big(\mathbf{\Psi}_2(p_1,q_1)\big)$ and $\mathbf{z}\big(\mathbf{\Psi}_2(p_2,q_2)\big)$, and select the larger one. The recursion continues until no ambiguity exists when making decisions.

\subsection{AFDM with Orthogonal Matching Pursuit (OMP)}
AFDM-IMI is an efficient heuristic method that exploits the partial information of relative amplitude, whereas AFDM-OMP makes use of more information in the derived CFR in (\ref{zyz}), thus achieving potentially better estimates. 
Specifically, although $\mathbf{\bar{H}}$ in (\ref{zyz}) has $N^2$ entries, it is determined by $P$ triplets of $\left(\mathbf{h}_P,\mathbf{a}_P,\boldsymbol{\tau}_P\right)$. Since the number of multipath components $P$ is typically smaller than the number of SCs $N$, it is possible that these $P$ paths can be identified by compressed sensing (CS) methods\cite{5895106} based on only a limited number of measurements (i.e., a limited number of pilot SCs $N_p\leq N$).
This problem can be solved by constructing a so-called \textit{complete dictionary}, made of the signals parameterized by a representative selection of possible parameters.
For our setup, we first rewrite $\mathbf{z}$ as


\begin{small}
\begin{equation}\label{strcture_signal}
\mathbf{z}=\big[\mathbf{\Gamma}\left(\tau_1,a_1\right)\mathbf{s},\cdots,\mathbf{\Gamma}\left(\tau_P,a_P\right)\mathbf{s}\big]\begin{bmatrix}
        h_1 \\ \vdots \\ h_P
    \end{bmatrix}+\bar{\mathbf{w}},
\end{equation}
\end{small}%
where $\mathbf{\Gamma}\left(\tau_i,a_i\right)\triangleq \mathbf{\bar{H}}_i, i=1,\cdots,P$. If the parameters $\left(\boldsymbol{\tau}_P,\mathbf{a}_P\right)$ were available, we could construct the $N\times P$-matrix in (\ref{strcture_signal}) and solve for $\mathbf{h}_P$ using least squares (LS).
Nevertheless, the exact parameters $\left(\boldsymbol{\tau}_P,\mathbf{a}_P\right)$ are not known a priori, and thus we choose representative sets of $\left(\boldsymbol{\tau}_P,\mathbf{a}_P\right)$ as

\begin{small}
\begin{equation}
    \tau\in\big\{0,\frac{1}{f_\mathrm{s}},\frac{2}{f_\mathrm{s}},\cdots,\tau_\mathrm{max}\big\},
\end{equation}
\begin{equation}
    a\in\left\{-a_\text{max},a_\text{max}+\Delta a,\cdots,a_\text{max}\right\}.
\end{equation}
\end{small}%

The discretization in $\tau$ is based on the assumption that after synchronization all arriving paths fall into the guard interval, where we choose the time resolution as the baseband sampling time $1/f_\mathrm{s}$, leading to $N_\tau=f_\mathrm{s}\tau_\mathrm{max}+1$ tentative delays.
On the other hand, assuming a maximum Doppler factor of $a_\text{max}$ and a resolution of $\Delta a$, we have $N_a=2a_\text{max}/\Delta a+1$ representative Doppler factors. Hence, a total of $N_\tau N_a$ candidate paths will be searched, and we expect $P\ll N_\tau N_a$ significant paths due to channel sparsity.
Then (\ref{strcture_signal}) can be rewritten as

\begin{small}
\begin{equation}\label{CompressedSensing}
\mathbf{z}=\left[\mathbf{\Gamma}\left(\tau_1,a_1\right)\mathbf{s},\cdots,\mathbf{\Gamma}\left(\tau_{N_\tau},a_{N_a}\right)\mathbf{s}\right]\begin{bmatrix}
        \xi_{1,1 }\\ \vdots \\ \xi_{N_\tau,N_a}
    \end{bmatrix}+\bar{\mathbf{w}}=\mathbf{\Phi}\boldsymbol{\xi}+\bar{\mathbf{w}},
\end{equation}
\end{small}%
where $\mathbf{\Phi}$ represents the complete dictionary which is a fat matrix with $N_\tau N_a$ columns, and $\boldsymbol{\xi}$ represents the complex channel gains to be estimated, whereby most of its entries are zero due to channel sparsity.
Note that the training vector $\mathbf{s}$ which only contains pilot SCs is known to the receiver side, and thus the dictionary $\mathbf{\Phi}$ could be computed.

After constructing the complete dictionary $\mathbf{\Phi}$ based on our derived AFDM input/output relations under MSML channels, we can then adopt off-the-shelf CS algorithms to solve the sparse estimation problem with the measurement model in
(\ref{CompressedSensing}). 
For illustration, we adopt Orthogonal Matching Pursuit (OMP)\cite{5895106}, a representative and efficient CS algorithm, which iteratively identifies one path at a time and solves a constrained LS problem at each iteration to measure the fitting error.
More details for the OMP procedure can be found in, e.g., \cite{li2007estimation}\cite{5895106}.


\subsection{Mutual Incoherence Property (MIP)}\label{SectionMIP}
\begin{figure}
	\centering
	\includegraphics[width=0.55\linewidth,  trim=0 0 0 10,clip]{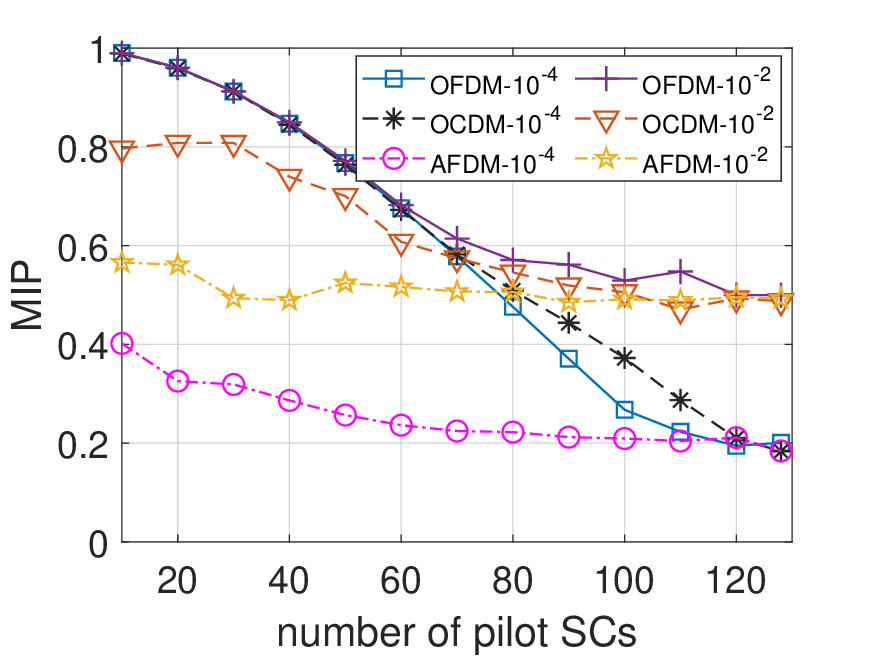}
	\caption{The MIP of OFDM-OMP, OCDM-OMP, and AFDM-OMP under MSML channels.\vspace{-2ex}}\label{Fig_OMP_MIP}
\end{figure}

A widely used metric associated with the quality of sparse signal recovery is the MIP\cite{5895106} defined as
\begin{small}
\begin{equation}\label{Formula_MIP}
    \mu\triangleq\max_{i\neq j}\frac{\left\|\langle \boldsymbol{\phi}_i,\boldsymbol{\phi}_j\rangle\right\|}{\left\|\boldsymbol{\phi}_i\right\|\cdot\left\|\boldsymbol{\phi}_j\right\|},%
\end{equation}%
\end{small}%
which represents the largest correlations among any two different columns $\boldsymbol{\phi}_i$ and $\boldsymbol{\phi}_j$ in the complete dictionary $\mathbf{\Phi}$. 
A smaller MIP suggests that any two multipath components are less similar with each other, and hence easier to be distinguished/estimated.
For illustration, the MIP for dictionaries of OFDM, OCDM and AFDM with different $N_p$ and under different Doppler factors is shown in Fig. \ref{Fig_OMP_MIP}, respectively.
First, it is observed that a higher number of pilot SCs helps reducing the MIP.
Second, as discussed in Section \ref{InputOutput}, higher Doppler factors spread the energy of each path to other paths, which increases their mutual correlations and hence the MIP, as can be seen in Fig. \ref{Fig_OMP_MIP}.
Nevertheless, AFDM still possesses the lowest MIP compared with OCDM and OFDM, due to its better MPS.


\section{Numerical Results}\label{SectionSimulation}
Numerical results are provided to evaluate the normalized mean square error (NMSE) of channel estimation and also BER performance of AFDM in MSML channels. The $h_i$s are generated as i.i.d. CSCG random variables, whose variance $\sigma_i^2$ (or power) decays exponentially with delay, with $\sum_i^P \sigma_i^2=1$. 
The following parameters are used if not mentioned otherwise: $N=128$, $P=5$, $N_a=3$, $l_\text{max}=19$, $Q_\text{max}=1$, $f_\mathrm{s}=1.5$ kHz, $f_\mathrm{c}=35$ kHz and Doppler factors in $10^{-4}$.


\subsection{OFDM, OCDM and AFDM under OMP Estimation}
\begin{figure}
        \centering
        \begin{subfigure}{0.5\linewidth}
                \centerline{\includegraphics[width=0.95\linewidth,  trim=0 0 25 0,clip]{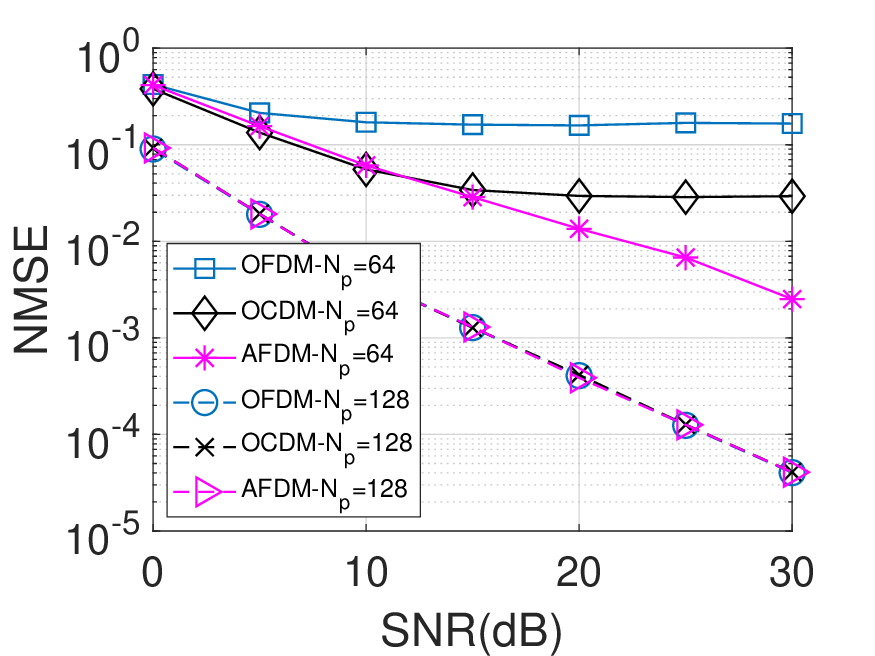}}
                \caption{}
                \label{Fig_OMP_NMSE}
        \end{subfigure}%
        \centering    
        \begin{subfigure}{0.5\linewidth}
                \centerline{\includegraphics[width=0.95\linewidth,  trim=0 0 25 0,clip]{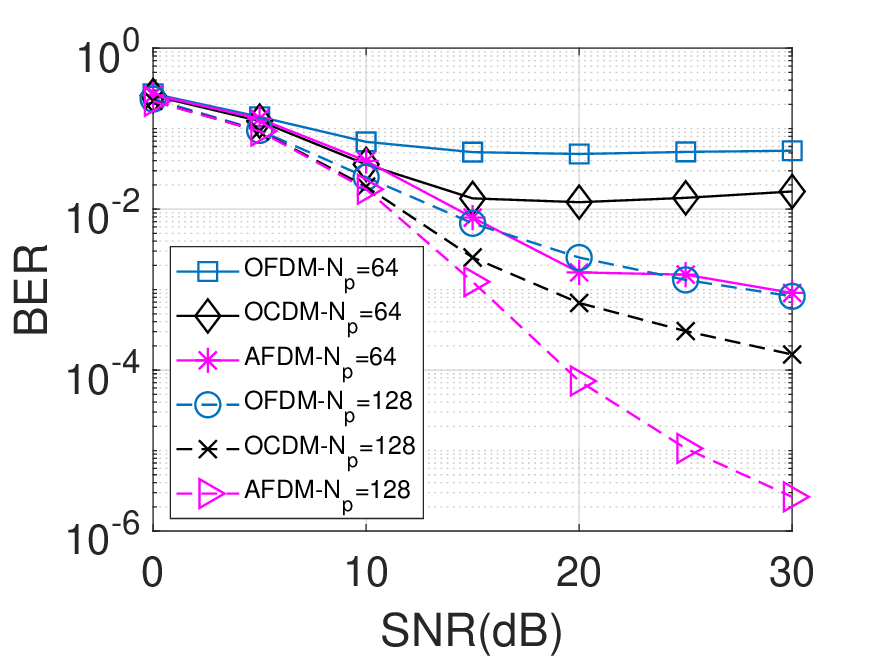}}
                \caption{}
                \label{Fig_OMP_BER}
        \end{subfigure}%
        \caption{The performance indicators of OFDM-OMP, OCDM-OMP, and AFDM-OMP under MSML channels: (a) NMSE (b) BER.\vspace{-2ex}}\label{Fig_results_OMP}
\end{figure}

Since AFDM includes OFDM and OCDM as special cases, here we first compare their performance under the same framework of OMP channel estimation. 
Fig. \ref{Fig_OMP_NMSE} and Fig. \ref{Fig_OMP_BER} show their NMSE and BER performance, respectively, with different number $N_p$ of pilot SCs out of total $N$ SCs in the pilot symbol. 
With a sufficient number of pilot SCs (e.g., $N_p=N=128$), it can be seen that AFDM-OMP, OCDM-OMP, and OFDM-OMP have almost the same channel estimation accuracy in terms of NMSE (effectively approaching the ideal CSI case), while their BER performance degrades in sequential order.
This could be attributed to the higher channel diversity gain of AFDM over OCDM and further OFDM, due to their different MPS capabilities, as discussed in Section \ref{SectionDiversity} and also in Fig. \ref{Fig_DO_PEP}.
On the other hand, with $N_p=64$, the NMSE performance deteriorates for all three schemes, while AFDM performs better than OCDM and further OFDM. 
Similar trends are observed for their BER performance.
The reduced number of pilot SCs causes the performance deterioration, while AFDM still performs the best out of the three schemes. 
The results could be jointly explained based on the COP metric in \eqref{Formula_COP} and the MIP metric in (\ref{Formula_MIP}).
As discussed in Sections \ref{SectionDiversity} and \ref{SectionMIP}, AFDM has the lowest COP and also lowest MIP, which jointly lead to lower NMSE and BER.

\subsection{Comparison of AFDM Channel Estimation Methods}
\begin{figure}
        \centering
        \begin{subfigure}{0.5\linewidth}
                \centerline{\includegraphics[width=0.95\linewidth,  trim=0 0 25 0,clip]{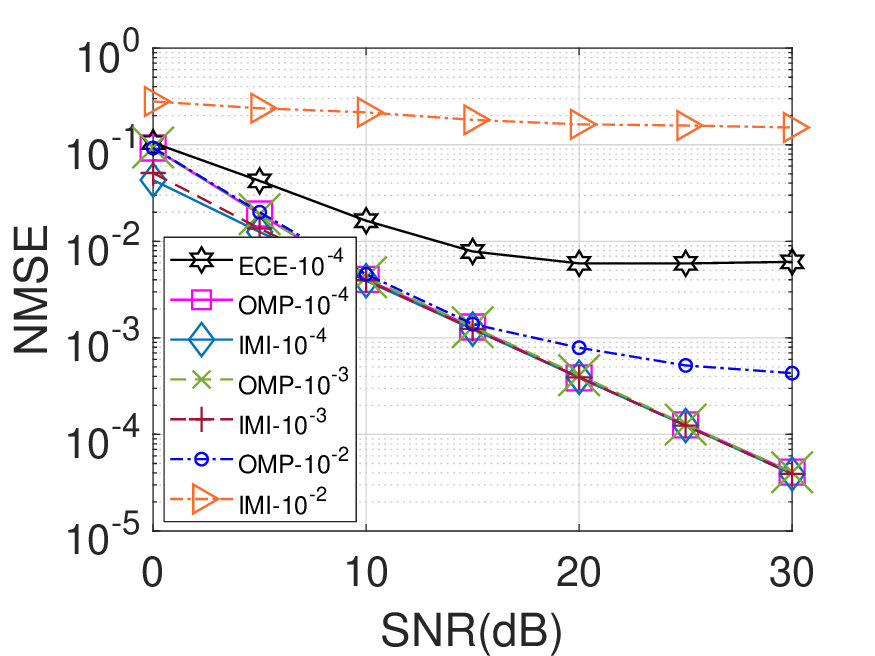}}
                \caption{}
                \label{Fig_AFDM_NMSE}
        \end{subfigure}%
        \centering    
        \begin{subfigure}{0.5\linewidth}
                \centerline{\includegraphics[width=0.95\linewidth,  trim=0 0 25 0,clip]{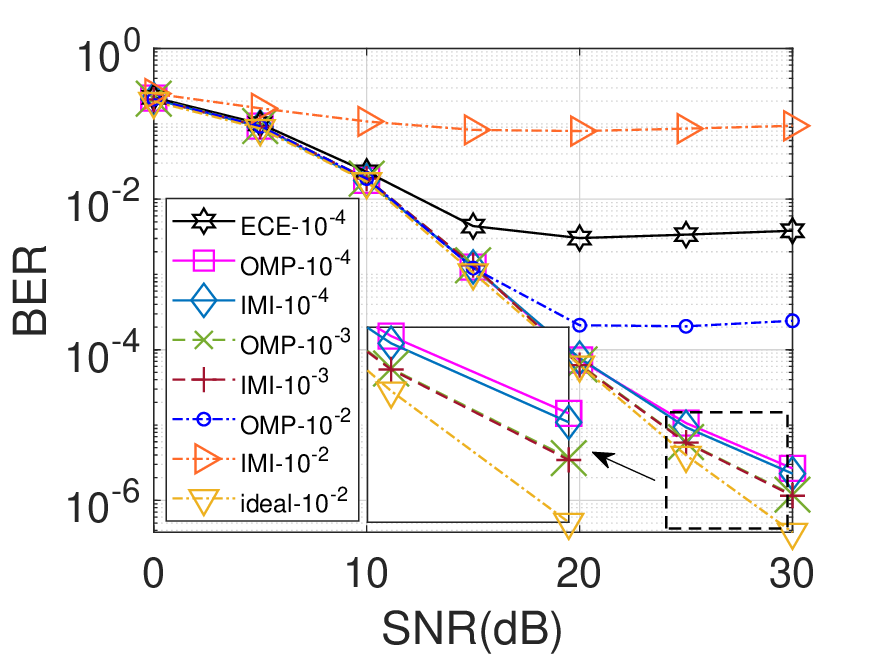}}
                \caption{}
                \label{Fig_AFDM_BER}
        \end{subfigure}%
        \caption{The performance indicators of AFDM-ECE, AFDM-IMI, and AFDM-OMP under MSML channels: (a) NMSE (b) BER.\vspace{-2ex}}\label{Fig_results_AFDM}
\end{figure}

Here we compare the NMSE and BER performance of AFDM-ECE, AFDM-IMI and AFDM-OMP, as shown in Fig. \ref{Fig_AFDM_NMSE} and Fig. \ref{Fig_AFDM_BER}, respectively.
Three channel configurations are considered with different levels of Doppler factors in the order of $10^{-4}$, $10^{-3}$ and $10^{-2}$, which correspond to a maximum movement speed of $1.8$ km/h, $18$ km/h and $41$ km/h, respectively.\footnote{CFs are set accordingly such that the DFSs are on-grid, whereas the case with off-grid DFSs is left for future work.}
Note that the original AFDM-ECE is not applicable under high Doppler factors (e.g., $10^{-3}$ and $10^{-2}$), due to ambiguity in separating the paths in MSML channels.
It can be seen that AFDM-IMI and AFDM-OMP have similar performance under low to moderate Doppler factors (e.g., $10^{-4}$ and $10^{-3}$), both outperforming the original AFDM-ECE.
On the other hand, under high Doppler factors (e.g., $10^{-2}$), the performance of AFDM-IMI deteriorates more significantly compared with AFDM-OMP, due to the increased ambiguity in the multi-paths. 
In contrast, AFDM-OMP better exploits our derived AFDM input/output relations under MSML channels, and achieves robust performance at the cost of slightly higher computational complexity.
In addition, it is interesting to observe that a higher Doppler factor (e.g., $10^{-3}$ versus $10^{-4}$) is not necessarily detrimental (e.g., see the enlarged window in Fig. \ref{Fig_AFDM_BER}), 
which might be attributed to the DTS diversity inherent in MSML channels, and is worth further explorations.
Finally, for ADFM-OMP, a gap from the BER lower bound with ideal CSI is still observed under high Doppler factors, which calls for even better channel estimation methods.




\section{Conclusions}
This paper introduces the new AFDM waveform to the complex UWA MSML channels, and addresses the challenging channel estimation problem where each multipath might have different Doppler scale in addition to Doppler shift.
Based on the newly derived input-output formula of AFDM and its characteristics, two new channel estimation methods are proposed, i.e., AFDM-IMI under low to moderate DTS, and AFDM-OMP (with slightly increased complexity) under high DTS. Numerical results confirm the effectiveness of the proposed methods against the original AFDM-ECE method. Moreover, the resulted AFDM system outperforms OFDM as well as OCDM in terms of channel estimation accuracy and BER performance, which could be attributed to the lower COP and MIP of multi-path observation dictionary in AFDM and hence also its higher channel diversity gain under MSML channels.
Future work may consider off-grid delays/Doppler shifts, performance evaluations in practical UWA environments and comparison with OTFS.
\bibliography{IEEEabrv,bibliography_AFDM}
\end{document}